\documentclass[twocolumn,preprintnumbers,superscriptaddress,prl,longbibliography,amsmath,amssymb]{revtex4-1}

\usepackage{graphicx}
\usepackage{dcolumn}
\usepackage{bm}
\usepackage{color}
\usepackage{ulem}
\usepackage{gensymb}
\usepackage{braket}
\usepackage{amsmath}
\usepackage[percent]{overpic}

\begin{document}

\title{Self-selecting vapor growth of transition metal halide single crystals}

\author{J.-Q. Yan}
\email{yanj@ornl.gov}
\affiliation{Materials Science and Technology Division, Oak Ridge National Laboratory, Oak Ridge, Tennessee 37831, USA}

\author{M. A. McGuire}
\affiliation{Materials Science and Technology Division, Oak Ridge National Laboratory, Oak Ridge, Tennessee 37831, USA}

\date{\today}

\begin{abstract}
Transition metal halides can host a large variety of novel phenomena, such as magnetism in the monolayer limit, quantum spin liquid and spiral spin liquid states, and topological magnons and phonons. Sizeable high quality single crystals are necessary for investigations of magnetic and lattice excitations by, for example, inelastic neutron scattering. In this paper, we review a less well-known vapor transport technique, self-selecting vapor growth, and report our growths of transition metal halides using this technique. We report the growth and characterizations of sizable single crystals of RuCl$_3$, CrCl$_3$, Ru$_{1-x}$Cr$_x$Cl$_3$, and CrBr$_3$. In order to expedite the conversion of starting powder to single crystals, we modified the technique by cooling the growth ampoule through an appropriate temperature range. 
Our work shows that the self-selecting vapor transport technique can provide large single crystals of transition metal halides, demonstrating its potential for providing high quality single crystals of quantum materials.

\end{abstract}

\maketitle

The study of materials confined to two spatial dimensions is associated with many hallmark discoveries and phenomena in condensed matter systems.  The bulk of this experimental work is traditionally done using carefully grown and finely tuned thin films, but more recently cleavable, layered materials have provided a complementary a top-down approach. Ultrathin flakes can be peeled from crystals comprising neutral layers joined to one another through van der Waals (vdW) bonding. This enables not only the study of 2D physics in single materials, but also the construction of vdW heterostructures combining functionalities\cite{geim2013van}. Transition metal halides represent one important class of such vdW layered compounds\cite{mcguire2017crystal}. These compounds often adopt structures that are relatively simple, yet contain interesting structural motifs like triangular or honeycomb networks. These materials appear over a wide breadth of current materials physics contexts, including 2D magnetic order\cite{huang2017layer}, multiferroicity\cite{kurumaji2013magnetoelectric,wu2012magnetic,kurumaji2011magnetic,tokunaga2011multiferroicity}, quantum spin liquids\cite{mcguire2019chemical,plumb2014alpha}, spiral spin liquids\cite{gao2022spiral}, topological magnons, and chiral phonons\cite{chen2018topological,schneeloch2022gapless, do2022gaps,yin2021chiral}. The availability of sizable, high-quality crystals of layered halides is important in enabling continued progress in these areas. 

Often transition metal halides can be grown by simple vapor transport or sublimation reactions \cite{BinnewiesGlaumSchmidtSchmidt+2012, may2020practical} due to their high vapor pressure at elevated temperatures. This typically produces platelike crystals with lateral dimensions from 1-10 mm, but with thicknesses limited to between 0.001 and, in rare cases, 0.1 mm. Some examples melt congruently, and in those cases Bridgman growths can be a better approach\cite{gao2022spiral}. While many 2D and device physics research applications require only small crystals to exfoliate, building the prerequisite detailed understanding of these materials often demands measurements on thicker crystals with larger mass and smaller aspect ratios. This is particularly true for neutron scattering techniques, which provide valuable information about the crystal and magnetic structures as well as lattice and magnetic excitations. 

In this work, we review the principles of a less well-known vapor transport technique, self-selecting vapor growth (SSVG)\cite{szczerbakow2005self}, and report the growth and characterization of single crystals of transition metal halides. This technique was developed about 50 years ago but has only been employed to grow II-VI and IV-VI semiconductors. We grew sizeable single crystals of transition metal halides using SSVG and found cooling through a temperature range is rather effective in expediting the conversion of the starting powder to crystals. This growth effort was motivated by the need of large single crystals to study the fractionalized excitations in quantum spin liquid candidate $\alpha$-RuCl$_3$ and bosonic Dirac physics in chromium halides (Cr\textit{X}$_3$, \textit{X}=Cr, Br, and I) with ferromagnetic honeycomb layers. For all transition metal halides mentioned in this work, conventional vapor transport growths using a large temperature gradient are used when small crystals are more appropriate for the planned experimental investigations.

\begin{figure*} \centering \includegraphics [width = 0.95\textwidth] {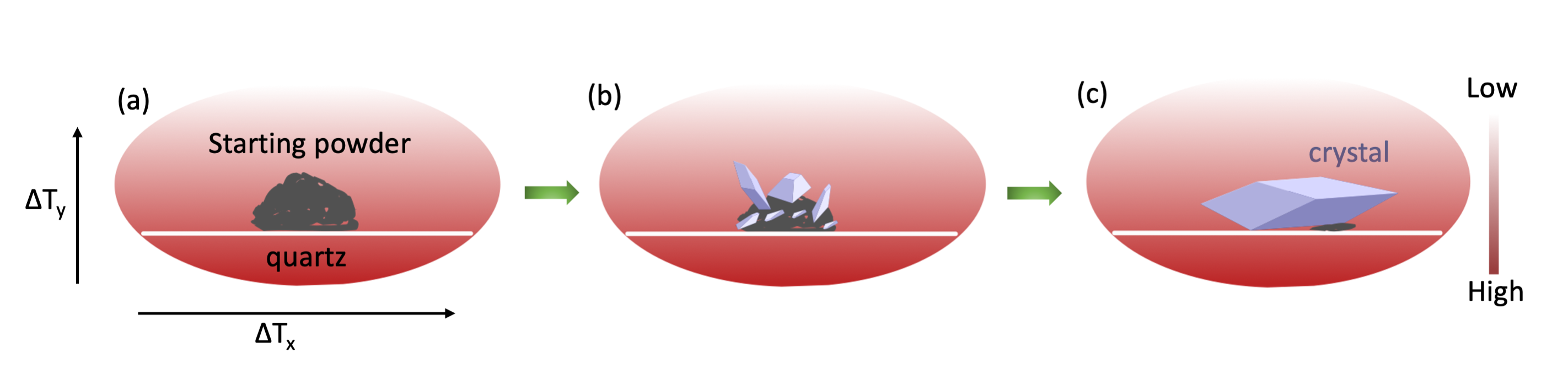}
\caption{(color online) Illustration of the self-selecting vapor transport growth performed inside of a sealed amouple, normally a fused-quartz tube. (a) starting powder in a well defined temperature gradient. Only a small portion of the quartz tube is drawn for simplification. A small temperature gradient is applied along the vertical direction as illustrated by the color scale. A small horizontal temperature gradient can be also applied to help the grain selection. (b) Over time, part of the starting powder is converted to small crystals growing  on top of the powder. (c) At the end of an ideal growth, one large single crystal forms consuming all starting powder and smaller crystals formed at the early stage of the growth.}
\label{illustration-1}
\end{figure*}

\section{Self selecting vapor transport growth}

Crystal growth by vapor transport makes use of the vapor phase generated from chemical reaction between the starting materials and the transport agent in a chemical vapor transport (CVT) growth, or from sublimation or decomposition of starting materials in a physical vapor transport (PVT) growth. PVT growths can occur through different routes: sublimation, decomposition sublimation, and auto transport\cite{BinnewiesGlaumSchmidtSchmidt+2012}. In a sublimation growth, the solid at the source end sublimates and the vapor deposits at the sink end of the growth ampoule. In the case of a decomposition sublimation process, the solid at the source end decomposes into gaseous phases, then crystals with the composition of initial solid can precipitate at the sink end out of the gaseous phases. In the so-called auto transport growth, the gaseous phases resulting from the decomposition of the initial solid act as the transport agent. The PVT growth mechanism of each transition metal halide varies depending on the composition, partial pressure, and role of the gaseous phases inside of the growth ampoule. More than one of these  mechanisms may occur in a single growth, for example, for the PVT growth of CrCl$_3$\cite{BinnewiesGlaumSchmidtSchmidt+2012}.

For most CVT and PVT growths, a large temperature gradient over, for example, 20$\degree$C is applied along the growth ampoule. This temperature gradient drives the mass transport that leads to the crystallization of the desired phase at the sink end. One natural result of this process is that the source powder and the resulting crystals stay at different ends of the growth ampoule and are separated from each other. This approach has been successfully employed to grow high quality single crystals of various materials including intermetallics, oxides, transition metal chalcogenides and halides\cite{BinnewiesGlaumSchmidtSchmidt+2012}.

SSVG is one special vapor transport growth that occurs in a much smaller temperature gradient with the crystals growing on top of the starting powder. The principles and applicability of this technique are well reviewed previously\cite{szczerbakow2005self}. Figure\,\ref{illustration-1} illustrates the principles of a SSVG. The loosely packed starting powder was placed in a small (for example, less than 2$\degree$C) but well controlled temperature gradient illustrated in Fig.\,\ref{illustration-1}(a). SSVG can be performed in a horizontal configuration, which can have both horizontal and vertical temperature gradients,  or a vertical configuration, which normally has a vertical temperature gradient only. As time goes by, crystals show up on top of the starting powder which is the coolest part of the powder. The term "self-selecting" comes from the fact that growth of crystals on the cool top of the polycrystalline source favors the fastest growing orientations. Crystals continue to form and grow with the vapor transport of the hot source materials. Over time, one grows at the expense of the starting powder and smaller grains formed at the early stage of the growth; eventually one large single crystal is obtained (see Fig.\,\ref{illustration-1}(c)) in an ideal growth.

Compared to other vapor transport growths, SSVG is distinguished with the following features: (1) The small temperature gradient makes the growth is nearly isothermal. This leads to exceptional compositional uniformity for solid solutions and is critical for systems where the distribution coefficient is temperature dependent. (2) Crystals grow on top of the starting powder which acts as a seed. There is no contact between the crystals and the growth ampoule or crucible. Thus the crystals are free of strain from the contact with the container.

SSVG can be a CVT process when transport agents are present or a PVT process otherwise. This technique has been employed to grow large crystals of semiconducting II-VI and IV-VI material systems. To the best of our knowledge, it hasn't been employed to grow other novel quantum materials. In this paper, we report the growth of large single crystals of RuCl$_3$, CrCl$_3$, Ru$_{1-x}$Cr$_x$Cl$_3$, and CrBr$_3$ using SSVG. Previous growths of semiconducting II-VI and IV-VI materials were performed at a fixed setup temperature and the growths can be time consuming. In order to expedite the self selecting vapor transport growth of transition metal halides, we modified the growths by cooling in an appropriate temperature range at a reasonably slow cooling rate which is more efficient in converting all the starting powder to one single crystal.

\begin{figure*} \centering \includegraphics [width = 0.75\textwidth] {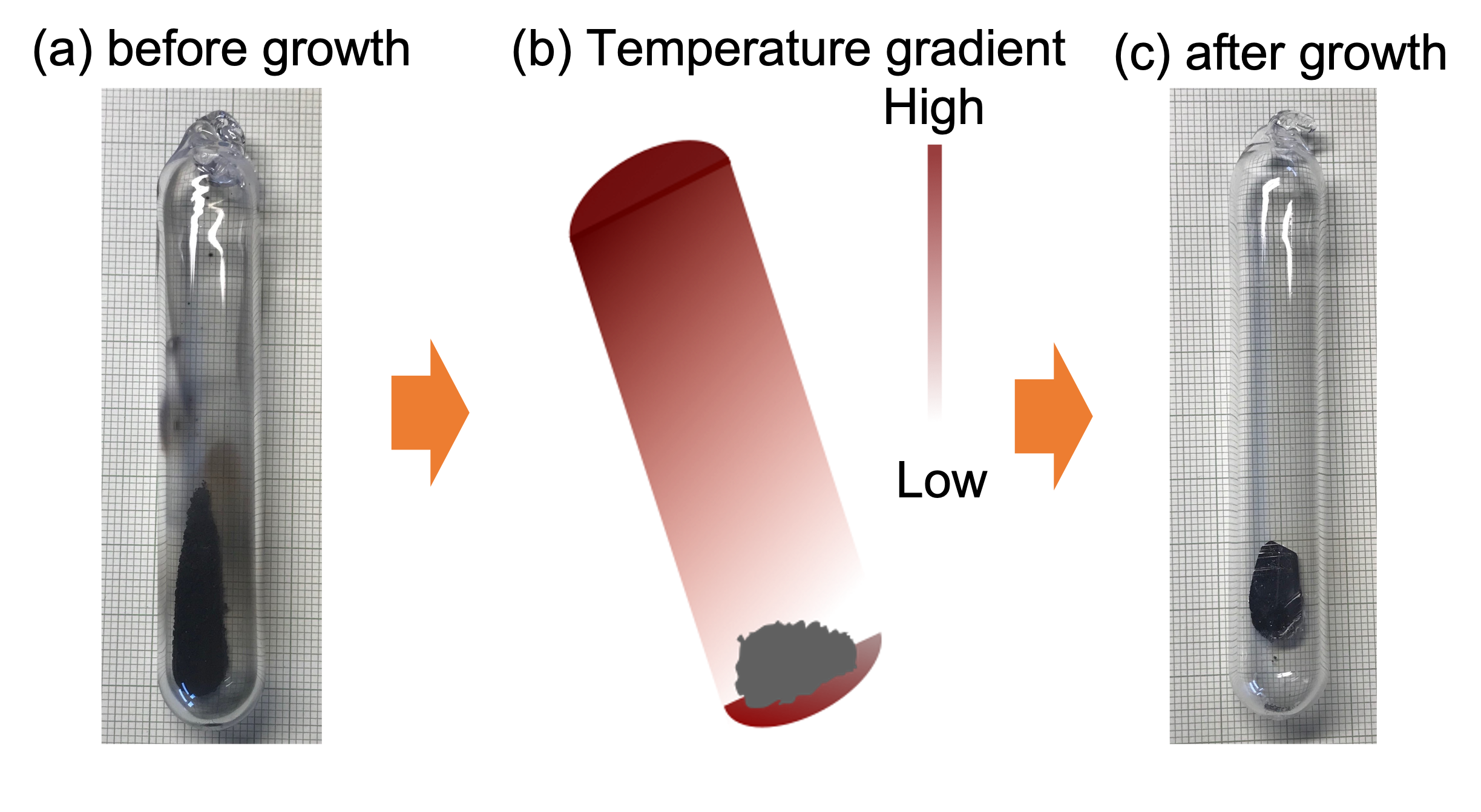}
\caption{(color online) Self-selecting vapor transport growth of RuCl$_3$. (a) about 1 g of RuCl$_3$ powder sealed in a fused quartz tube under vacuum. (b) The temperature gradient configuration used in our growths of all transition metal halides mentioned in this work. Note both the vertical and lateral components. The ampoule can be tilted to any angle as needed in addition to horizontal or vertical positions (see Fig.\,\ref{Furnace-1}). (c) All starting powder is converted to one single crystal after growth. As discussed in the text, details of the temperature gradient and cooling rate control the shape and dimension of the resulting crystals.}
\label{sublimation-1}
\end{figure*}

\begin{figure*} \centering \includegraphics [width = 0.95\textwidth] {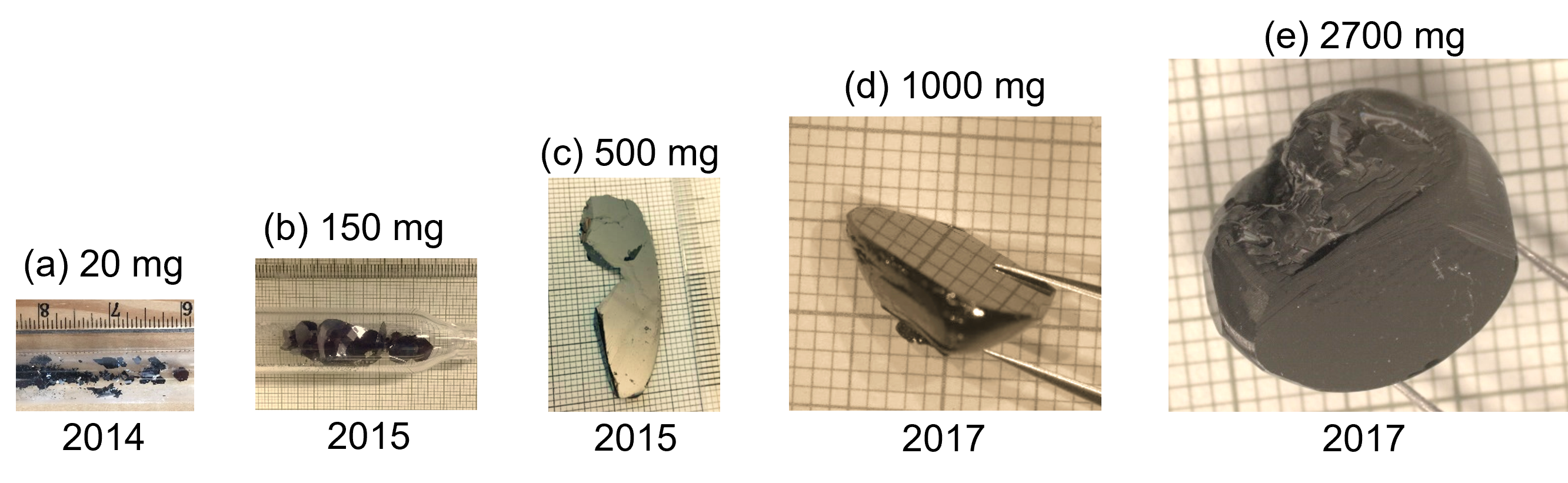}
\caption{(color online) $\alpha$-RuCl$_3$ crystals grown at different stages of our work. (a) A conventional vapor transport growth  performed in a horizontal tube furnace with a temperature gradient of about 45$\degree$C between the starting powder at the hot end and crystals at the cold end. (b-e) Self-selecting vapor transport growths performed in a box furnace. The increasing crystal dimension results from a better control of the temperature gradient. The number beneath each panel shows in what year the growth was performed.  The picture in Panel (c) was taken by Arnab Banerjee. $\alpha$-RuCl$_3$  powder purified by Craig Bridges at ORNL was used for the growth of crystals in (a-c). Commercial $\alpha$-RuCl$_3$  powder from Furuya Metals was used to grow the large crystals shown in (d,e). }
\label{RuCl3-1}
\end{figure*}

\section{$\alpha$-RuCl$_3$}

$\alpha$-RuCl$_3$ continues to be intensely studied as a candidate Kitaev spin liquid\cite{takagi2019concept}. Inelastic neutron scattering is one of the few direct probes of the fractionalized excitations in quantum spin liquid materials, therefore, the availability of large single crystals is critical. This compound does not  melt but sublimes at high temperatures. This indicates that growths using solidification from a molten state, such as Bridgman, Czochralski, and floating zone techniques, are not applicable to the growth of $\alpha$-RuCl$_3$ crystals. Instead, vapor transport should be an effective approach. In Table I of Ref.[\citenum{kim2022alpha}], Kim \textit{et al.} summarized the vapor transport conditions reported in the literature. Single crystals of $\alpha$-RuCl$_3$ can be grown in horizontal tube furnaces with a purposely controlled temperature gradient with the aid of a transport agent such as Cl$_2$ (730-660$\degree$C in Ref.[\citenum{majumder2015anisotropic}], 750-650$\degree$C in Ref.[\citenum{hentrich2018unusual}]) or TeCl$_4$ (700-650$\degree$C in Ref.[\citenum{may2020practical}]) or even no transport agent since $\alpha$-RuCl$_3$ has a reasonable vapor pressure at elevated temperatures. In these types of growths, the starting $\alpha$-RuCl$_3$ powder is kept at the hotter end of the growth ampoule during the growth; plate-like $\alpha$-RuCl$_3$ crystals form at the cooler end after an extended stay at high temperatures. $\alpha$-RuCl$_3$ crystals grown in this manner are normally plates with limited thickness. 

We employ SSVG to grow $\alpha$-RuCl$_3$ single crystals that are typically larger than 1 gram per piece for inelastic neutron scattering (INS) measurements. These large single crystals with a small mosaic allows INS measurements without coaligning many smaller pieces. More importantly, by carefully controlling the temperature gradient around the powder, we can control the shape of resulting crystals as illustrated in Fig.\,\ref{RuCl3-1}(c) and (d). A slightly larger temperature gradient along the ampoule favors the growth of platelike crystals.

In the early stage of our work on this project, commercial $\alpha$-RuCl$_3$ powder from Alfa Aesar was first purified before crystal growth or further studies\cite{banerjee2016proximate}. In recent growths, two types of $\alpha$-RuCl$_3$ powder are used. Most growths were performed using the commercial powder from Furuya Metals (Japan). Some growths were performed using $\alpha$-RuCl$_3$ powder obtained from AlCl$_3$-KCl salt. The detailed procedure for producing $\alpha$-RuCl$_3$ powder by reacting RuO$_2$ and AlCl$_3$-KCl salt was reported previously\cite{yan2017flux}. X-ray powder diffraction measurements found no impurity phases in either powder samples. Magnetic measurements show a magnetic ordering temperature of 14\,K for the commercial powder and 12\,K for the powder made from the salt flux. In the whole temperature range 2\,K-300\,K, the commercial powder has a larger magnetization. 

The starting powder was sealed under dynamic vacuum in a fused quartz ampoule. For the growth of crystals about 1\,g/piece, the fused quartz tube used has an outer diameter of 19\,mm, a wall thickness of 1.5\,mm, and a length of approximately 100\,mm. A tube with an outer diameter of 25\,mm was used instead when the growth used 2\,g or more of material. The sealed ampoule was then put inside of a box furnace with a temperature gradient as shown in Fig.\,\ref{sublimation-1}(b). After dwelling at a furnace set point temperature of 1060$\degree$C for 6h, the furnace was cooled to 800$\degree$C at 4$\degree$C/h. Power to the furnace was then powered off to cool to room temperature. When the cooling rate is in the range 2-4$\degree$C/h, all of the powder inside of the ampoule converts into one single crystal. As shown in Fig.\,\ref{RuCl3-1}(d,e), the crystals are rather thick and look like Chinese style Go stones. Platelike crystals can be obtained by increasing the cooling rates or by controlling the temperature gradient along the quartz ampoule. These large crystals have minimal amount of stacking faults\cite{cao2016low} and have been used in INS experiments by our collaborators\cite{banerjee2018excitations, banerjee2017neutron}. 

Kubota et al. reported the growth of $\alpha$-RuCl$_3$ crystals using a Bridgman furnace\cite{kubota2015successive}. In this growth, the growth ampoule was pulled downward in a Bridgman furnace at a rate of 3\,mm/h over 80 hours. This resembles the operation for a typical Bridgman growth out of melt. However, it should be noted that only solid and vapor phases are involved in the growth of $\alpha$-RuCl$_3$ based on our understanding of the growth mechanism. This growth performed inside of a Bridgman furnace is still one type of vapor transport growth. In such a growth, the starting powder can be located at any position inside of the growth ampoule, often at the top or the bottom end. Regardless of the position of the starting powder, as long as the crystals are found on top of the powder or where the powder is in case of a complete conversion of starting powder to crystals, this growth inside of a Bridgman growth is also a self-selecting vapor transport growth. In this case, the conversion of powder to crystal is controlled by the vertical temperature profile and the pulling rate. However, if the crystals grow at a different and cooler position of the ampoule and are separated from the starting powder, this growth can be understood as vapor transport growths performed inside of a tube furnace with a larger temperature gradient. The growth temperature gradient can be one important factor determining the crystal quality and properties. The recently discussed sample dependence of thermal transport properties indicates that the growth conditions can have a dramatic effect on the stoichiometry, layer stacking sequence, and structural defects, and hence the physical properties\cite{lee2021}.

\section{CrCl$_3$}
\begin{figure} \centering \includegraphics [width = 0.47\textwidth] {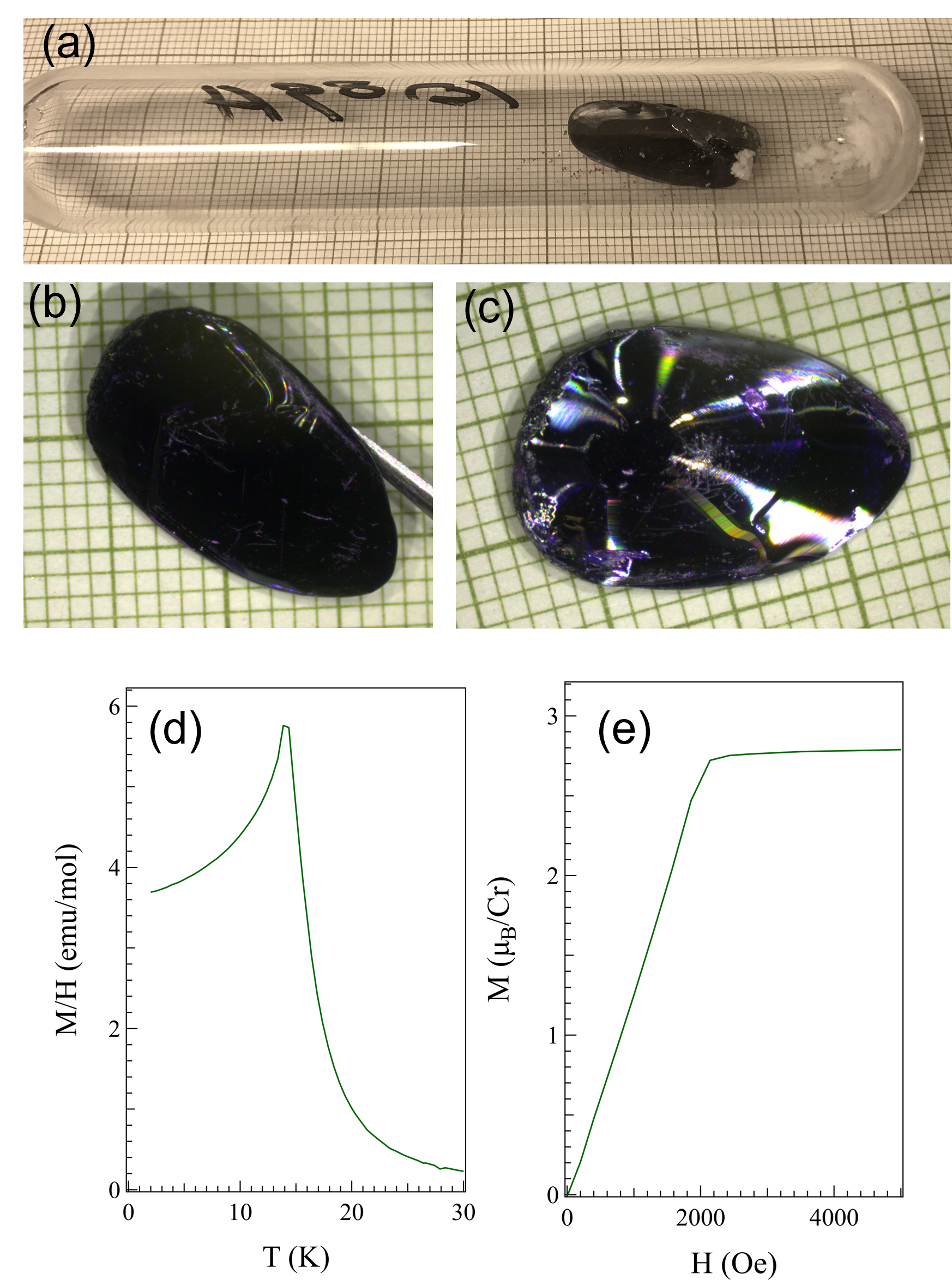}
\caption{(color online) (a) Picture of one CrCl$_3$ crystal of 1.2\,g inside of the growth ampoule. The white fluffy material at the bottom of the ampoule is amorphous SiO$_2$ resulting from the reaction between residual moisture, CrCl$_3$, and quartz tube as described in the text. (b,c) A closer view of both sides of the Go-stone-like crystal. (d) Temperature dependence of magnetization in a magnetic field of 100\,Oe applied along the ab plane. (e) In-plane field dependence of magnetization at 2\,K. }
\label{CrCl3Mag-1}
\end{figure}

Cr trihalides (Cr\textit{X}$_3$, \textit{X}=F, Cl, Br, I) with ferromagnetic honeycomb layers are model systems to study the magnonic Dirac physics\cite{pershoguba2018dirac}. Large single crystals are needed to study the bosonic excitations by, for example, INS. In order to grow high quality CrCl$_3$ crystals, we first purified commercial CrCl$_3$ by vapor transport technique. The commercial CrCl$_3$  was first dried in dynamic vacuum overnight at room temperature, then sealed in a fused quartz tube with an outer diameter of 19\,mm, a wall thickness of 1.5\,mm, and a length of 150\,mm. The sealed ampoule was put inside of a tube furnace with the end with dry CrCl$_3$  at 700$\degree$C and the cold end at 400$\degree$C. A large amount of magenta-violet transparent CrCl$_3$  crystals were found at the cold end after 12 hours.  About 1g of these small crystals were then transferred into a new quartz tube of the same diameter but only 100\,mm long. The sealed ampoule was put inside of a box furnace and heated to 950$\degree$C, held at this temperature for 2 hours and then cooled down to 600$\degree$C at 3$\degree$C/h. The furnace was then turned off to cool to room temperature. Figure\,\ref{CrCl3Mag-1}(a) shows the growth ampoule after growth. Occasionally some white residue can be found at the bottom of the growth ampoule. Elemental analysis and x-ray powder diffraction measurements confirm this is amorphous silica resulting from the reaction of residual moisture, CrCl$_3$, and the quartz tube. In the extreme situation, green colored Cr$_2$O$_3$ might show up inside of the growth ampoule, which is a good indication of water contamination of the starting materials. Using well dried starting materials as described above would significantly reduce the amount of these secondary phases in the products. Just like the growth of RuCl$_3$ described above, CrCl$_3$ crystals grown in this manner can be rather thick with a shape of one piece of Go stone. Figure\,\ref{CrCl3Mag-1}(b,c) shows a closer view of both sides of the crystal. Magnetic measurements (see Figure\,\ref{CrCl3Mag-1}(d,e)) confirm CrCl$_3$ crystals grown by SSVG show the same magnetic properties as those thin crystals grown by conventional vapor transport technique\cite{mcguire2017magnetic}. The temperature dependence of magnetization in the paramagnetic state (not shown) shows a structure transition around 250\,K as reported previously\cite{mcguire2017magnetic}. 

One piece of CrCl$_3$ crystal of 0.88 gram grown by SSVG technique was used in an INS study performed by our collaborators\cite{do2022gaps}. The mosaic of this crystal is 0.68 degree, much smaller than that of coaligned crystals which is normally several degrees. The small mosaic turns out to be important in accurately determine the gaps in topological magnon spectra.

\section{CrBr$_3$}

\begin{figure} \centering \includegraphics [width = 0.47\textwidth] {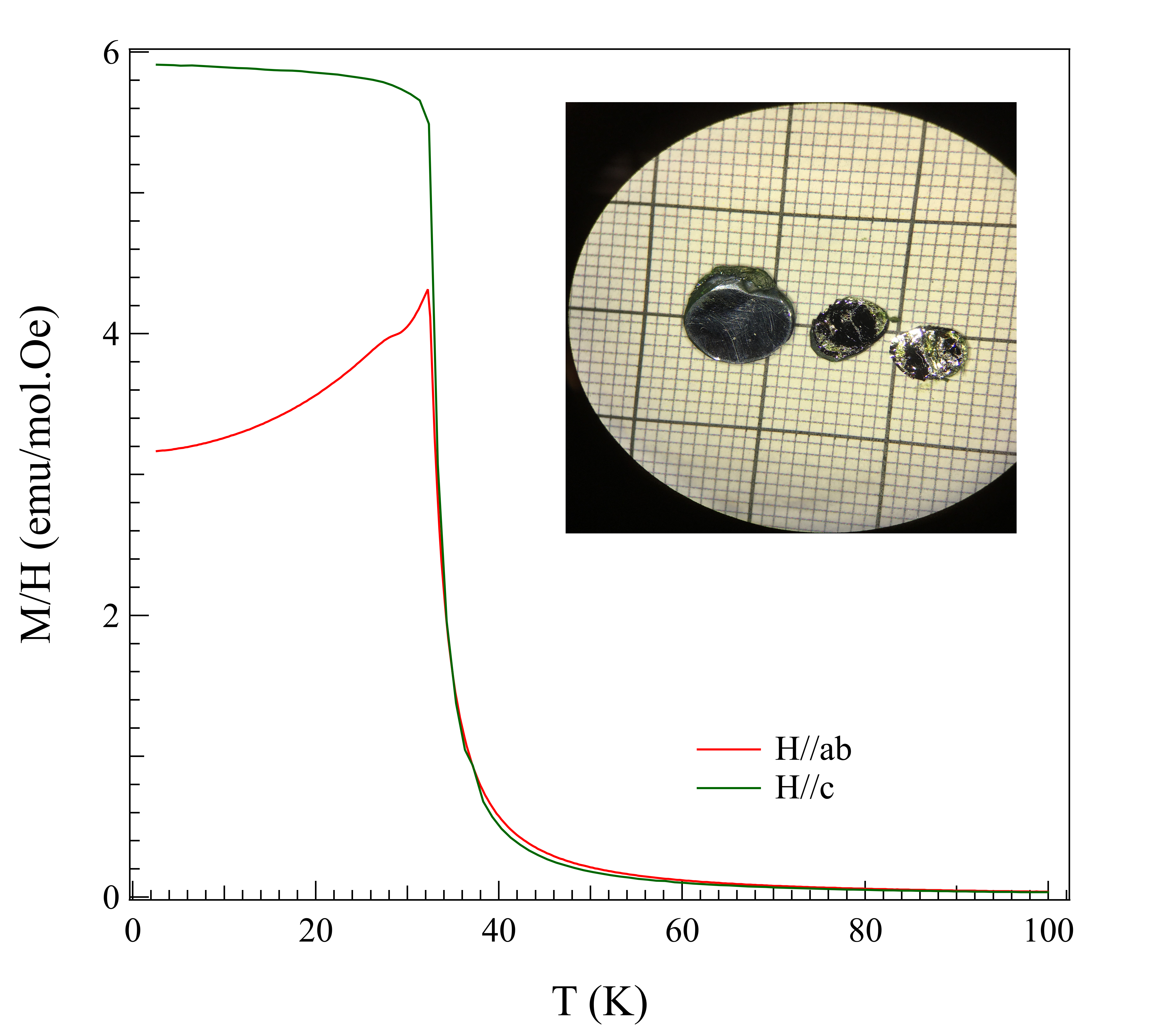}
\caption{(color online)  Temperature dependence of magnetization measured in a field cool mode using an applied magnetic field of 100\,Oe. Inset shows a photograph of the CrBr$_3$ crystals, with the smaller of the two cleaved into two (center and right). }
\label{CrBr3-1}
\end{figure}

CrBr$_3$ powder used for the crystal growth was synthesized by reacting Cr powder with bromine produced by decomposition of CuBr$_2$. One of us developed this double chamber reaction ampoule specifically for this and similar growths. The details for this synthesis were reported previously\cite{may2020practical}. About 0.8\,g of CrBr$_3$ powder was sealed in a fused quartz tube with an outer diameter of 19\,mm, a wall thickness of 1.5\,mm, and a length of 100\,mm.  The sealed ampoule was put inside of a box furnace and heated to 950$\degree$C, held at this temperature for 6 hours and then cooled down to 600$\degree$C at 2$\degree$C/h. The furnace was then powered off to cool to room temperature. This procedure enables a full conversion of all starting powder into two single crystals (along with some green residue). The inset of Fig.\,\ref{CrBr3-1} shows the intact larger crystal and the smaller crystal cleaved in two. Figure\,\ref{CrBr3-1} shows the anisotropic temperature dependence of magnetization measured in an applied magnetic field of 100\,Oe. The magnetic ordering temperature is 33\,K, consistent with previous results\cite{yu2019large}.

\section{Ru$_{1-x}$Cr$_x$Cl$_3$}
\begin{figure} \centering \includegraphics [width = 0.47\textwidth] {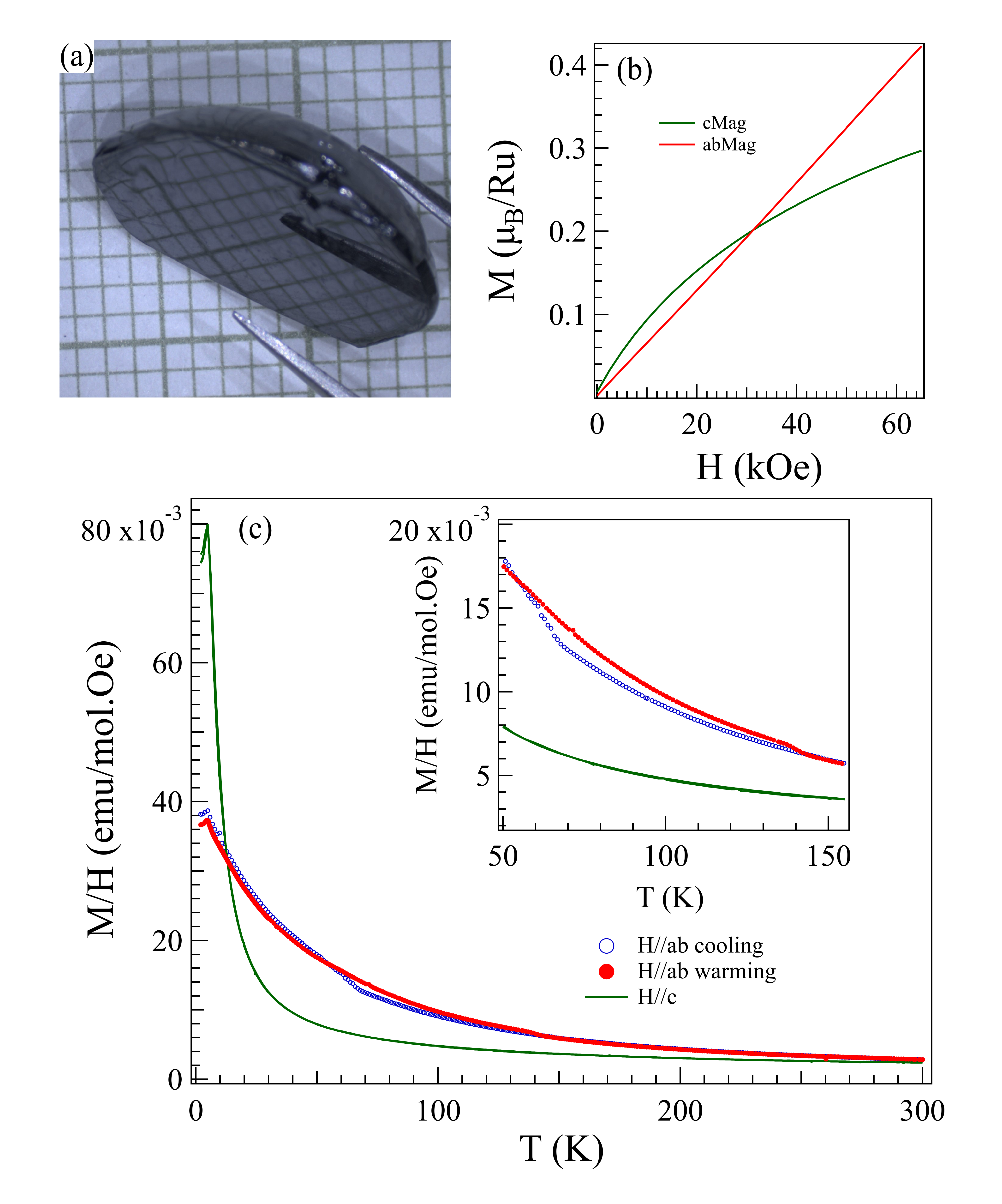}
\caption{(color online)  (a) Ru$_{1-x}$Cr$_x$Cl$_3$ crystals grown by self-selecting vapor transport. (b) Field dependence of magnetization at 2\,K. (c) Temperature dependence of magnetic susceptibility measured in an applied magnetic field of 1\,kOe. Inset highlights the loop induced by the first order structural transition.}
\label{RuCrCl3-1}
\end{figure}

In addition to high pressure and high magnetic field, chemical substitution\cite{bastien2019spin, do2018short,lampen2017destabilization} is an effective approach to manipulating the magnetic ground state and competing interactions in $\alpha$-RuCl$_3$. The growth of large single crystals of doped $\alpha$-RuCl$_3$ is rather challenging mainly because the dopant can have a quite different vapor pressure than the parent compound. Here we use 
Ru$_{1-x}$Cr$_x$Cl$_3$ as an example to demonstrate that large single crystals of uniform solid solutions of transition metal halides can be obtained by SSVG possibly due to the small temperature gradient. To grow Cr-substituted $\alpha$-RuCl$_3$, sublimation cleaned small CrCl$_3$  crystals and well dried $\alpha$-RuCl$_3$ powder obtained from AlCl$_3$-KCl salt were used as the starting materials. The same growth parameters as those for $\alpha$-RuCl$_3$ were used for the growth of lightly doped compositions. Figure\,\ref{RuCrCl3-1}(a) shows the picture of one Ru$_{0.9}$Cr$_{0.1}$Cl$_3$ crystal. A portion of the crystal was carefully removed and used for the magnetic measurements and elemental analyses. The Ru/Cr ratio was determined by energy dispersive spectroscopy on different cleaved surfaces. We found no significant variation in Cr concentration laterally across cleaved surfaces or through the thickness of the crystal comparing different cleaved surfaces. Figure\,\ref{RuCrCl3-1}(b) shows the field dependence of magnetization measured with the field applied perpendicular and parallel to the plate. A linear field dependence is observed when the magnetic field is applied parallel to the plate. This is in contrast to a nonlinear field dependence when the field is perpendicular to the plate. Figure\,\ref{RuCrCl3-1}(c) shows the temperature dependence of magnetization. With 10\% of Ru substituted by Cr, the magnetic ordering temperature is suppressed from 7\,K in RuCl$_3$ to near 5\,K in Ru$_{0.9}$Cr$_{0.1}$Cl$_3$. We observed only one anomaly around 5\,K which indicates the absence of stacking faults\cite{cao2016low}. The doping dependence of magnetic ordering temperature agrees with previous reports.\cite{bastien2019spin, roslova2019detuning}  Thermally hysteretic behavior is seen in  the temperature range 70\,K-140\,K when the magnetic field is parallel to the plate, likely associated with the first order structural phase transition\cite{cao2016low}. Compared to that in $\alpha$-RuCl$_3$, the structural transition occurs in a wider temperature range. The structure transition in CrCl$_3$ occurs around 250\,K, which is about 100\,K higher than that in $\alpha$-RuCl$_3$. It would be interesting to monitor how the structure transition evolves with the substitution in Ru$_{1-x}$Cr$_x$Cl$_3$.

\section{Discussion}

\begin{figure} \centering \includegraphics [width = 0.47\textwidth] {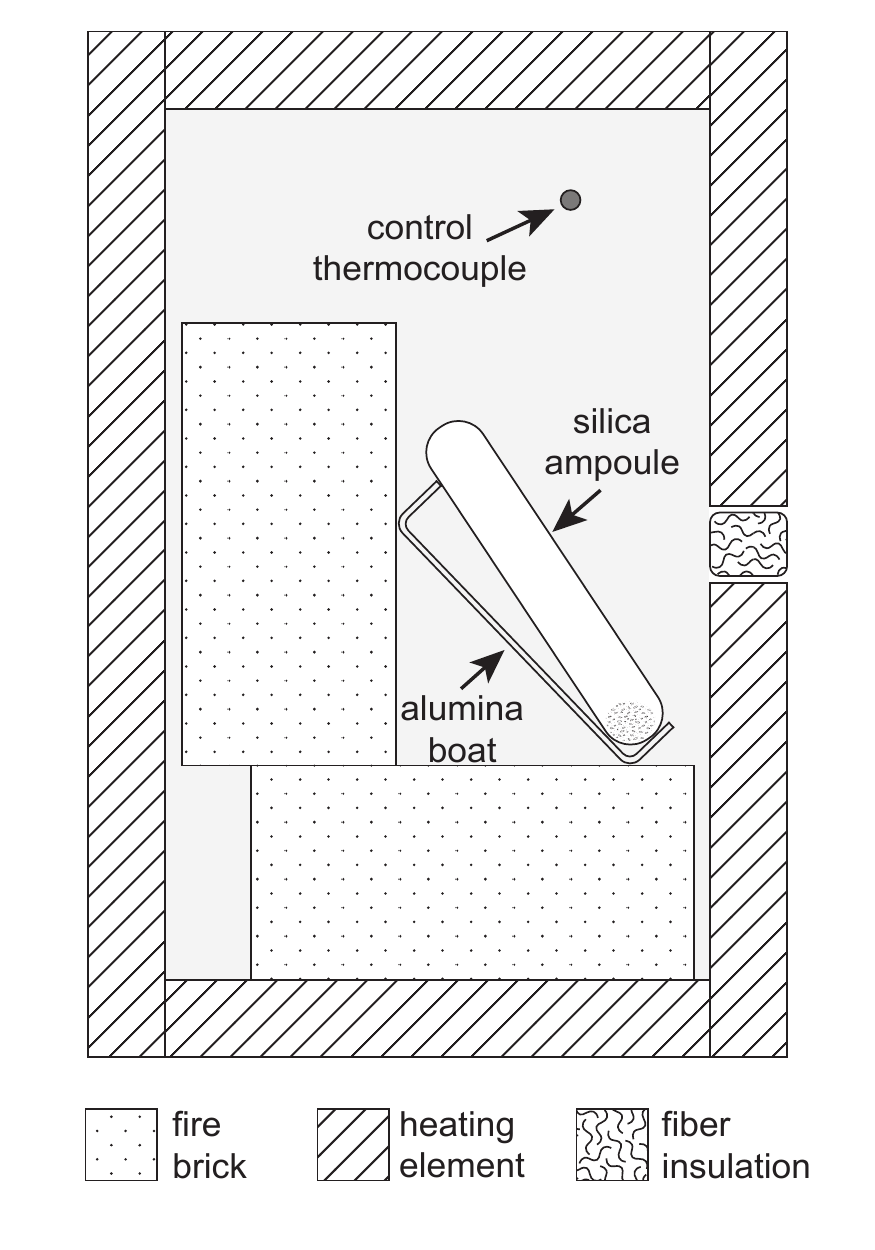}
\caption{(color online)  Schematic picture of the furnace that we used to grow all transition metal halides and chalcogenides mentioned in this work. The amount of fiber insulation inside of the ventilation hole helps create the desired temperature gradient near the bottom end of the growth ampoule.}
\label{Furnace-1}
\end{figure}

One unique feature of SSVG is the unusually small temperature gradient applied vertically and/or horizontally to the starting powder. However, in reality, it is rather challenging to obtain such a small but well defined temperature gradient. The numbers beneath crystal pictures in Fig.\,\ref{sublimation-1} show how long it takes for us to gradually get a good control of the temperature gradient ideal for the growth of large RuCl$_3$ crystals even though we realized its importance at the early stage of this project. Such a small temperature gradient can be obtained in either a tube furnace or a box furnace. We tested the SSVGs using tube furnaces and obtained some beautiful crystals. However, we found that it is harder to control the number and position of the nucleation sites when a tube furnace, either one-zone or multi-zone, is used. This is likely due to variation of the detailed temperature profile of each furnace. We thus performed the growths in a box furnace making good use of the natural temperature gradient of the furnace. The temperature inside of a box furnace is not uniform and it is rather challenging to measure the temperature variation at different positions inside of the furnace. The consequence of this is that we do not know the exact temperature gradient near the powder. However, it is expected to be rather small. Figure\,\ref{Furnace-1} shows how Al$_2$O$_3$ fire bricks and fiber insulation are employed inside of a box furnace to create a desired temperature gradient near the bottom of the growth ampoule. The amount of fiber insulation inside of the ventilation hole can be adjusted for this purpose. We identified the position best for SSVG by positioning the growth ampoules at different  locations along the fire bricks. We noticed that crystals always appear at the coldest place of the ampoule, which is a good indicator of the detailed temperature profile. When there is a temperature gradient along the ampoule and the starting powder stays at a hot place, crystals formed away from the powder often have a few large grains. In case of the absence of any temperature gradient around the starting powder, many pieces of small platelike crystals are obtained. Occasionally, we obtained nice crystals by keeping the growth ampoule upright inside of an alumina crucible. In this case, the horizontal temperature gradient is negligible and mainly the vertical temperature gradient drives the growth.

Cooling at an appropriate rate obviously promotes the transformation from powder to crystals in a SSVG. We tried the growth of $\alpha$-RuCl$_3$ at 1000$\degree$C and 900$\degree$C in the temperature gradient optimized for SSVG. The crystallization is not complete even after a month. After confirming the importance of cooling during the crystallization, we further tested how the cooling rate affects the shape and size of crystals. All test growths were performed using a 1g batch. If the cooling rate is above 8$\degree$C/h, some powder remains inside of the growth ampoule and several pieces of crystals are normally found. One single crystal is normally observed if the cooling rate is lower than 4$\degree$C/h. For growths with a cooling rate in the range of 4-8$\degree$C/h, either one piece of crystal with large in-plane dimension or a few pieces with some smaller in-plane dimension typically result. The above knowledge learned from the growth of $\alpha$-RuCl$_3$ worked well for other transition metal halides mentioned in this work. We thus choose different cooling rates depending on the crystal geometry desired for specific measurements. Changes in the crystal growth resulting from small changes in the placement of the ampoule in the furnaces suggest the detailed temperature gradient near the powder has a dominant effect on the shape of the grown crystals. 

Two experimental observations indicate that the total pressure inside of the growth ampoule affects the nucleation and growth. We tried SSVG of MoCl$_3$, CrI$_3$, and Os$_{0.55}$Cl$_2$ but the conversion of power to crystal is never complete and we obtained only millimeter sized single crystals even with  a slow cooling rate of 1$\degree$/h \cite{mcguire2019chemical, mcguire2017high}. All three compounds have a high vapor pressure even at temperatures below 500$\degree$C. The high vapor pressure might promote nucleation. An even longer stay at relatively low temperatures might be desired. The other experimental observation indicating the detrimental effects of high pressure is that high purity starting materials facilitate the growth of large single crystals. Some commercial chemicals are not pure and may have volatile impurities. Full conversion of starting powder to crystals is normally observed after purifying the commercial starting materials. This motivates us to purify, for example, CrCl$_3$, before crystal growth as described above. Also for the same reason, in some growths of $\alpha$-RuCl$_3$, we used high purity powders synthesized by reacting RuO$_2$ with AlCl$_3$-KCl\cite{yan2017flux}. For materials with high vapor pressure the growth temperature should be kept relatively low, which may also require a slower cooling rate to allow complete reaction.

SSVG might be employed to grow quantum materials more than transition metal halides presented in this work and II-VI and IV-VI semiconductors reported in literature. Recently, vapor transport growth in a small temperature gradient was employed to grow intrinsic antiferromagnetic topological insulator MnBi$_2$Te$_4$ and related compounds\cite{yan2022vapor,hu2021growth}. Motivated by this, we tried the SSVG of MnBi$_2$Te$_4$. Sub-millimeter sized crystalline plates are obtained on top of the starting powder after dwelling at 565$\degree$C for over three weeks. More growths and characterizations are in progress to investigate whether SSVG can provide a better control of the lattice defects, which are believed to play an essential role in realizing the elusive quantum anomalous Hall effect in MnBi$_2$Te$_4$\cite{yan2022perspective}, than flux growth\cite{yan2019crystal} and chemical vapor transport\cite{yan2022vapor}. We also tested the SSVG of MoSe$_2$ and WSe$_2$ in the presence of I$_2$  and obtained millimeter sized crystals by keeping the growth ampoule at 1025$\degree$C for over a month. These preliminary growths suggest that SSVG can be employed to grow other quantum materials.  

\section{Summary}

In summary, we review the self-selecting vapor growth technique. This technique was developed over half century ago, but its application has been limited to the II-VI and IV-VI semiconductors. We report the growth of large single crystals of transition metal halides using this technique. We performed the SSVG by cooling through a temperature range which has been proved to expedite the conversion of the starting powder to crystals. The temperature range ideal for the growth is determined by the vapor pressure of the starting materials. An ideal vapor pressure should be high enough to allow efficient material conversion but slow enough  to control the number of resulting crystals. A slower cooling rate is favored for a complete conversion of the starting powder to one single large crystal. Preliminary growths on transition metal chalcogenides suggest that this technique works well for other quantum materials.

\section{Acknowledgment}
The authors would thank collaborations and discussions with Arnab Banerjee, Craig Bridges, Huibo Cao, Andrew Christianson, Seunghwan Do, Shang Gao, Paige Kelley, David Mandrus, Andrew May, Stephen Nagler, Brian Sales, and Alan Tennant. This work was supported by the US Department of Energy, Office of Science, Basic Energy Sciences, Materials Sciences and Engineering Division. Crystal growth of RuCl$_3$ after 2020 was supported by the U.S. Department of Energy, Office of Science, National Quantum Information Science Research Centers, Quantum Science Center. 

 This manuscript has been authored by UT-Battelle, LLC, under Contract No.
DE-AC0500OR22725 with the U.S. Department of Energy. The United States
Government retains and the publisher, by accepting the article for publication,
acknowledges that the United States Government retains a non-exclusive, paid-up,
irrevocable, world-wide license to publish or reproduce the published form of this
manuscript, or allow others to do so, for the United States Government purposes.
The Department of Energy will provide public access to these results of federally
sponsored research in accordance with the DOE Public Access Plan (http://energy.gov/
downloads/doe-public-access-plan).

\section{references}
%

\end{document}